\documentclass[10pt,twocolumn]{article}
				\usepackage{osajnl2_plop}
				\usepackage{tabularx}
				\usepackage{subfigure}
				\usepackage[latin1]{inputenc}
				\usepackage[T1]{fontenc}
				\usepackage{amsmath,amsfonts,mathrsfs}
				\usepackage{flushend}
				\graphicspath{{EPS/}}
				
\begin{document}
				\twocolumn[
				\title{Off-axis digital hologram reconstruction: some practical considerations}
				
				\author{N. Verrier and M. Atlan}
				\address{ Institut Langevin. Fondation Pierre-Gilles de Gennes. Centre National de la Recherche Scientifique (CNRS) UMR 7587, Institut National de la Sant\'e et de la Recherche M\'edicale (INSERM) U 979, Universit\'e Pierre et Marie Curie (UPMC), Universit\'e Paris 7. \'Ecole Sup\'erieure de Physique et de Chimie Industrielles (ESPCI ParisTech) - 10 rue Vauquelin. 75005 Paris. France }
				
				\email{nicolas.verrier@espci.fr}
				
				\begin{abstract}
			Holographic rendering of off-axis intensity digital holograms is discussed. A review of some of the main numerical processing methods, based either on the Fourier transform interpretation of the propagation integral or on its linear system counterpart, is reported. Less common methods such as adjustable magnification reconstruction schemes or Fresnelet decomposition are presented and applied to the digital treatment of off-axis holograms. The influence of experimental parameters on the classical hologram reconstruction methods is assessed, offering guidelines for optimal image rendering regarding to the hologram recording conditions.
				\end{abstract}
				
				\ocis{090.1995, 100.3010}
				]
				
				\section{Introduction}
				\label{intro}
			Optical holography consists of the acquisition of images from diffracted optical field measurements. Holographic imaging was initially proposed by Gabor~\cite{Gabor1948} for electron microscopy. Holograms were recorded on high resolution photographic plates. Due to the subsequently-called "in-line" configuration, holograms were stained with twin-image and zero-order contributions, which overlapped with the signal image~\cite{Gabor1949}. Originally recorded with the red radiation of Mercury lamps, holograms were increasingly recorded with laser light sources, which gave much more reliable results. In 1962, Leith and Upatnieks proposed to introduce an off-axis reference beam~\cite{Leith1962} to separate, in the spatial frequency domain, the real image from the twin-image and zero-order diffraction terms. However, holograms were still to be reconstructed by optical means.\\
				
				The first digital reconstructions of optically-measured holograms were realized by Goodman~\cite{Goodman1967} and further by Kronrod~\cite{Kronrod1972} (in Russian. More details can be found in Ref.~\cite{Yaroslavsky}) in the early 1970's. Here, optically magnified parts of the holograms are digitally sampled and then reconstructed using Fourier-transform based routines. Digitalization of optical holograms allowed, for instance, to improve reconstruction quality~\cite{Onural1987}, to retrieve information about the phase of the recording wave~\cite{Fienup1982,Liu1987}, and to treat holograms without reconstruction~\cite{Onural1992}. One of the major breakthroughs in holographic imaging was initiated, by Schnars, with direct recording of digital holograms~\cite{Schnars1994}. Charge-Coupled Device (CCD) and Complementary Metal Oxide Semiconductor (CMOS) digital sensor arrays enabled the acquisition and numerical processing of high-resolution holograms at fast rates.\\
				
				Intrinsic properties of holographic imaging allow this technique to be used in a wide range of domains such as fluid mechanics~\cite{Lozano1999,Meng2004,Pu2005,Dubois2006,Atlan2006,Desse2008,Verrier2008a,Verrier2009,Verrier2010}, biomedical imaging~\cite{Schedin2000,Kim2000,XuW2001,Charriere2006,Kemper2008,Simonutti2010} or mechanical vibration analysis~\cite{Powell1965,Aleksoff1971,Zhang2004a,Picart2005,Leval2005,Borza2005,Iemma2006,Asundi2006,Joud2009}. Democratization of high resolution CCD and CMOS sensors played a major role in the development of digital reconstruction techniques. For instance, reconstruction of phase-only~\cite{Yamaguchi2006}, shifted~\cite{Matsushima2010}, tilted or aberrated data~\cite{Lebrun2003,Denicola2005,Verrier2008b} has been successfully demonstrated. Inverse-problem approaches make it possible to improve object localization and field of view in the reconstructed hologram~\cite{Soulez2007a,Soulez2007b}. Compressive sensing based approaches are also to be considered when working in noisy or low-light conditions~\cite{Denis2009,Marim2011}. Moreover, owing to the massive parallelization of image processing calculations by Graphics Processing Units (GPU), hologram reconstruction can be performed in real-time~\cite{ShimobabaSato2008,Ahrenberg2009,Shimobaba2010,SamsonVerpillat2011}.\\
				
				In this paper, we will describe most of the common off-axis digital holographic reconstruction schemes, and discuss their applicability. After some brief reminders about digital holographic recording, we will present the main reconstruction approaches, involving one to three Fourier transforms. Then a discussion about reconstruction with adjustable magnification is proposed. Methods to tackle aliases and replicas are proposed, leading to high quality magnified reconstructions. The use of Fresnelet transform will also be discussed. Reconstruction methods will be assessed experimentally with optically-acquired off-axis holograms, to provide insight into their respective suitability towards targeted applications.
				
				\section{Fresnel holography bases}
				\label{sec1}
				Digital holography typically consists of recording an optical field emerging from an illuminated object in a diffraction plane (\emph{e.g.} in free-space propagation conditions), and numerically calculating, from diffraction models, the field distribution in the reconstruction plane. In practice, optical holograms are measured-out from the interference of the diffracted beam beating against a reference beam, which is not disturbed by the object to be analyzed. One of the object-reference cross terms typically yields a complex-valued map (\emph{i.e.} quadrature-resolved : in amplitude and phase) of the diffraction field in the sensor plane. The complex-valued measurement contains relevant information about the local retardation of the diffracted field. Phase-shifting \cite{Yamaguchi1997} and frequency-shifting \cite{Atlan2007} techniques were proposed to record the diffraction field in quadrature. The interference pattern, recorded by sensor array, can be expressed as~\cite{Goodman}
			\begin{multline}
			 E\left(x,y\right)=\left|\mathcal{R}\left(x,y\right)\right|^2+\left|\mathcal{O}\left(x,y\right)\right|^2\\ +\mathcal{O}^*\left(x,y\right)\mathcal{R}\left(x,y\right)+\mathcal{O}\left(x,y\right)\mathcal{R}^*\left(x,y\right),
			\label{eq:RecCCD}
			\end{multline}
				where $\mathcal{R}$ and $\mathcal{O}$ denote reference and object optical fields respectively. Starred ($*$) symbols are associated with complex conjugate values.
			
				\begin{figure}[h]
				\centering
				\subfigure[]{\includegraphics*[width=4cm]{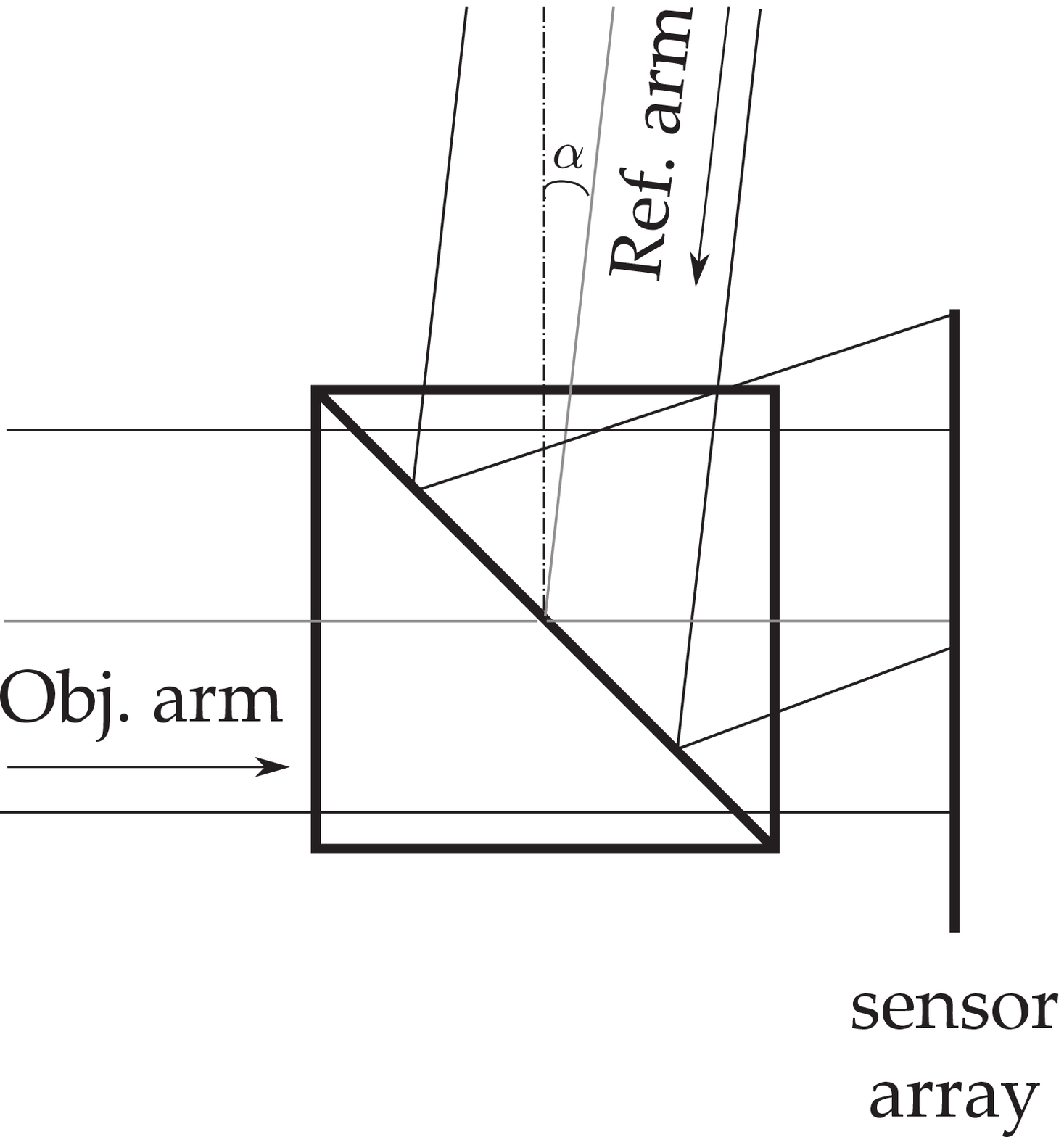}}
				\subfigure[]{{\includegraphics*[width=3.5cm]{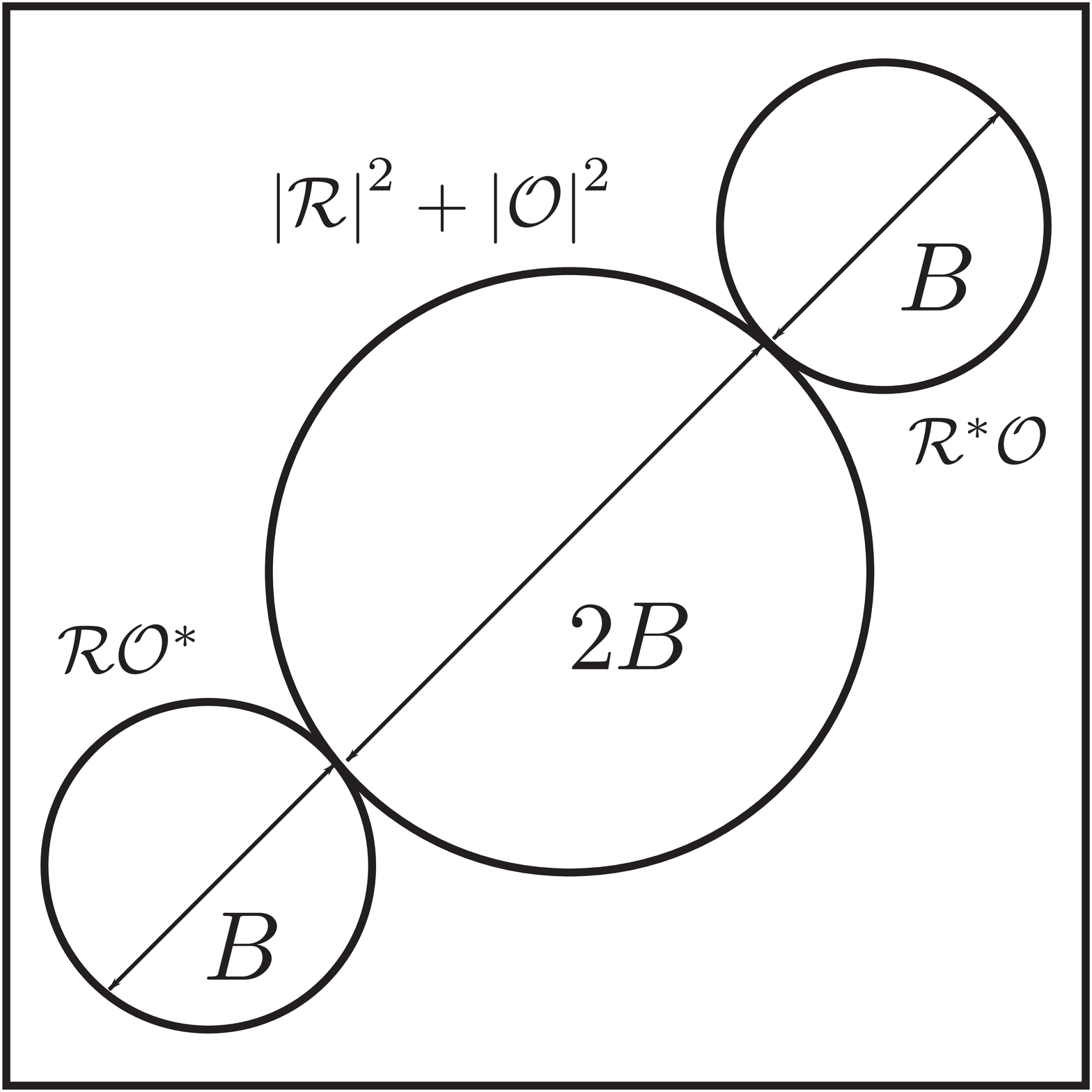}}}
				\caption{(a) Hologram recording in off-axis configuration. (b) Spatial frequency representation of off-axis holograms.}
				\label{fig1}
				\end{figure}				
				
			The interference between reference and object beams can be recorded within a wide range of configurations. These can be grouped in two main categories: in-line~\cite{Goodman} and off-axis configurations~\cite{Leith1963}. For our applications, off-axis holograms will be recorded with a Mach-Zehnder configuration. The final part of our off-axis Mach-Zehnder configuration is illustrated on Fig. (\ref{fig1}) (a). Here, reference and object beam are combined, with a relative angle $\alpha$, using a non-polarizing beam-splitting/combining cube. It should be noted that $\alpha$ should be chosen so as to fulfill the sampling theorem. The maximal value, leading to a correct sampling of the interference pattern is therefore given, under paraxial conditions, by:
			\begin{equation}
			\alpha_{\rm{max}}\approx\frac{\lambda}{2\Delta x},
			\end{equation}
			where $\Delta x$ denotes the sampling rate of the recording device.			
				
				This off-axis angle results in separation, of the four terms of Eq. (\ref{eq:RecCCD}), in the spatial frequency domain. This aspect is proposed Fig. (\ref{fig1}) (b). The central part of the hologram spectrum ($\left|\mathcal{R}\left(x,y\right)\right|^2+\left|\mathcal{O}\left(x,y\right)\right|^2$) is known as the autocorrelation term, its size is associated with the highest spatial frequencies of the object, denoted by $B$. Real and twin images of the object are respectively given by $\mathcal{O}\mathcal{R}^*$ and $\mathcal{R}\mathcal{O}^*$. These two terms are twice as small as the autocorrelation term. To improve the reconstruction quality, autocorrelation and twin image terms have to be canceled. This can be achieved either by spatial filtering~\cite{Cuche2000} or phase shifting~\cite{Yamaguchi2001}. Thus doing makes it possible to reconstruct the real image term only.
				
				In the following part, we will focus on hologram reconstruction. After a brief reminder about the Fresnel transform, we will discuss its main digital implementations.
				
				\section{Digital hologram reconstruction}
				\label{sec2}
				Digital reconstruction of an hologram consists in a \emph{a posteriori} refocusing over the original object, which can be performed by calculating backward propagation of the light from the hologram to the reconstruction plane. This process is equivalent to positioning the recorded hologram back into the reference beam. Reference beam therefore become the reconstruction beam.  Using the Huygens-Fresnel principle, one can infer an integral formulation of the intensity $E_{\rm{rec}}$ in the reconstruction plane, from an off axis recorded hologram $E$~\cite{Born}
			\begin{equation}
			 E_{\rm{rec}}\left(\xi,\eta\right)=-i\frac{z}{\lambda}\int_{\mathbb{R}^2}E\left(x,y\right)\frac{\exp\left(ikr\right)}{r}dx dy.
			\label{eq:HFresnel}
			\end{equation}
			Here, the distance $r$ is given by:
			\begin{equation}
			r=\sqrt{z^2+\left(x-\xi\right)^2+\left(y-\eta\right)^2},
			\end{equation}
				$\left(x,y\right)$ and $\left(\xi,\eta\right)$ denote the spatial coordinates in the hologram and reconstruction plane respectively. It should be noted that, the hologram $E\left(x,y\right)$ has been recorded  an off-axis configuration.
				Under Fresnel approximation, when $z^3>>\frac{1}{8\lambda}\left[\left(\xi-x\right)^2+\left(\eta-y\right)^2\right]^2$, $r$ can be approximated by:
			\begin{equation}
			 r=z\left[1+\frac{1}{2}\left(\frac{x-\xi}{z}\right)^2+\frac{1}{2}\left(\frac{y-\eta}{z}\right)^2\right],
			\end{equation}
			and Eq. (\ref{eq:HFresnel}) is rewritten as:
			\begin{multline}
			 E_{\rm{rec}}\left(\xi,\eta\right)=\frac{\exp\left(\frac{i2\pi}{\lambda}z\right)}{i\lambda z}\int_{\mathbb{R}^2}E\left(x,y\right)\\
			\times\exp\left\{i\frac{\pi}{\lambda z}\left[\left(x-\xi\right)^2+\left(y-\eta\right)^2\right]\right\}dx dy.
			\label{eq:FRT}
			\end{multline}
			This relationship will be used in the remainder of this paper to perform reconstruction of off-axis intensity holograms.
				
				As far as variables in Eq. (\ref{eq:FRT}) are separable, all the discrete formulations will be derived in the 1D case. Generalization in two dimensions is straightforward.
				The 1D discrete Fresnel transform is defined by:
			\begin{multline}
			E_{\rm{rec}}\left(p\right)=\frac{\exp\left(\frac{i2\pi}{\lambda}z\right)}{i\lambda z}\exp\left(i\frac{\pi}{\lambda z}p^2\Delta\xi^2 \right)\sum_{n=0}^{N-1}E\left(n\right)\\
			\times\exp\left(i\frac{\pi}{\lambda z}n^2\Delta x^2 \right)\exp\left(-i\frac{2\pi}{\lambda z}n p\Delta x \Delta\xi \right),
			\label{eq:DFRT}
			\end{multline}
			where $n\Delta x$ and $p\Delta\xi$ respectively denote the spatial coordinate in the CCD and reconstruction plane, and N is the number of sampling points.
			
			Direct implementation of Eq. (\ref{eq:DFRT}) is a time consuming process. Starting from Eq. (\ref{eq:FRT}), one can realize that efficient computational schemes can be designed to implement digital holographic reconstruction. This makes it possible to separate reconstruction methods into two main families: the Fourier based approaches (based on the use of a single fast Fourier transform (FFT)) well suited for imaging extended objects localized far from the CCD or CMOS sensor, and the convolution methods, computed by using two or three FFTs. These methods are well adapted for the reconstruction of holograms, of small lateral dimensions, recorded near the imaging device. Alternative methods can be considered when an adjustable magnification or advanced filtering techniques are needed.
				
				In the remainder of this section, we will detail the different computational approaches and apply these to the reconstruction of digital holograms.
				
				\subsection{Single-FFT method}
			Efficient implementation of Eq. (\ref{eq:DFRT}) can be performed using FFT algorithm~\cite{Cooley1965,Schnars2002}. In this case pixel pitches in both reconstruction ($\Delta\xi$) and CCD plane ($\Delta x$) are related by:
			\begin{equation}
			\Delta \xi=\frac{\lambda z}{N\Delta x}.
			\label{eq:g_fresnel}
			\end{equation}
			Therefore, Eq. (\ref{eq:DFRT}) can be rewritten as:
			\begin{multline}
			E_{\rm{rec}}\left(p\right)=\frac{\exp\left(\frac{i2\pi}{\lambda}z\right)}{i\lambda z}\exp\left(i\frac{\pi\lambda zp^2}{N^2\Delta x^2}\right)\\
			\times\sum_{n=0}^{N-1}E\left(n\right)\exp\left(i\frac{\pi}{\lambda z}n^2\Delta x^2 \right)\exp\left(-i2\pi\frac{np}{N}\right).
			\label{eq:1TF}
			\end{multline}
			This relationship is therefore easily computed by
			\begin{multline}
			E_{\rm{rec}}\left(\xi\right)=\frac{\exp\left(\frac{i2\pi}{\lambda}z\right)}{i\lambda z}\exp\left(i\frac{\pi\lambda zp^2}{N^2\Delta x^2}\right)\\
			\times\mathcal{F}\left\{E\left(x\right) \exp\left(i\frac{\pi}{\lambda z}x^2\right)\right\}.
			\end{multline}
			It should be noted that, within this configuration, the ratio between the reconstructed horizon and the sensor array extension (which can be abusively denoted as the magnification of the reconstruction method) is closely linked to the reconstruction distance \emph{i.e.} $\gamma=\Delta\xi / \Delta x=\lambda z /(N\Delta x^2)$. In the remainder of this paper, the intrinsic magnification of the single-FFT implementation of the reconstruction integral will be denoted by $\gamma_0=\lambda z /(N\Delta x^2)$.
				
				\subsection{Convolution based approaches}
Holographic reconstruction can be viewed as a linear system. As matter of fact, Eq. (\ref{eq:FRT}) is the mathematical expression of the spatial convolution between the hologram, and the Fresnel impulse response function $h_{z}$, which is defined by (omitting the multiplicative constant):
			\begin{equation}
			h_{z}\left(x\right)=\exp\left(i\frac{\pi}{\lambda z}x^2\right).
			\label{eq:Rep_imp}
			\end{equation}
			Convolution based approaches lead to unitary magnification, namely $\Delta \xi = \Delta x$.
				
				\subsubsection{``Three-FFT algorithm''}
			Computation of the convolution product between the hologram and the holographic impulse response can be efficiently implemented in Fourier domain. Using fast Fourier transform algorithms, Eq. (\ref{eq:FRT}) can be computed as:
			\begin{equation}
			 E_{\rm{rec}}\left(\xi\right)=\frac{\exp\left(\frac{i2\pi}{\lambda}z\right)}{i\lambda z}
			 \mathcal{F}^{-1}\left[\mathcal{F}\{E\left(x\right)\}\mathcal{F}\{h_{z}\left(x\right)\}\right],
			\end{equation}
			where $\mathcal{F}$ and $\mathcal{F}^{-1}$ respectively stand for the Fourier transform and its inverse.	
				
				\subsubsection{Angular spectrum propagation}
			This method is based on the propagation of the angular spectrum of the hologram. The angular spectrum transfer function is given by~\cite{LYu2005}:
			\begin{equation}
			H\left(u\right)\approx\exp\left[2i\frac{\pi z}{\lambda}\left(1-\frac{1}{2}\lambda^2u^2\right)\right],
			\label{eq:AngSpec}
			\end{equation}
			where $u$ is the spatial frequency in Fourier domain.
			Using Eq. (\ref{eq:AngSpec}), hologram reconstruction can be performed:
			\begin{equation}
			 E_{\rm{rec}}\left(\xi\right)=\frac{1}{i\lambda z}
			\mathcal{F}^{-1}\left[\mathcal{F}\{E\left(x\right)\}H\left(u\right)\right],
			\end{equation}

				\begin{figure}[h]
				\centering
				\includegraphics*[width=7.5cm]{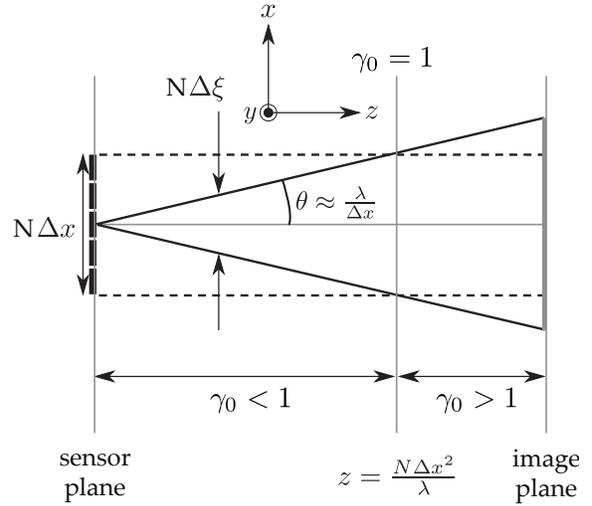}
				\caption{Angular acceptance of digital holographic reconstruction process. Solid lines are associated with the 1-FFT reconstruction, and dashed lines correspond to the convolution approaches.}\label{fig2}
				\end{figure}

				\subsection{Algorithms with adjustable magnification}
				Neither the single-FFT approach nor the convolution based methods allow the adjustment the magnification of the reconstructed hologram. As a matter of fact, in a single-FFT hologram processing, magnification depends on the recording wavelength and distance, whereas it remains constant using a convolution approach. In the latter case, the magnification is unitary ($\Delta \xi = \Delta x$). This aspect is illustrated by fig. (\ref{fig2}). Here, the evolution of the reconstructed horizon is represented with respect to the reconstruction distance. The solid lines are associated with the 1-FFT reconstruction scheme, and the doted lines are the reconstruction horizon of the convolution based reconstruction approaches. It should be noted that for $z=N\Delta x^2/\lambda$ (this distance is determined by taking $\Delta x=\Delta \xi$ in Eq. (\ref{eq:g_fresnel})), 1-FFT and convolution approaches exhibit the same magnification.
				
				Working with an adjustable magnification algorithm is a great opportunity to make the reconstruction horizon independent from hologram recording parameters. Domains such as multi-wavelength holography benefit from this property~\cite{JCLi2009,Picart2009}.

			Several approaches have been proposed to allow magnification adjustment. Ferraro used zero-padding to control the reconstructed horizon and to make it independent from the reconstructed distance~\cite{Ferraro2004}. This method gives good results for multiwavelength hologram multiplexing, but may however, increase the computational load. Another way to adjust the magnification is to reconstruct the hologram with a two step algorithm~\cite{Zhang2004}. Each step consists of a 1-FFT reconstruction. Let $z$ be the reconstruction distance, the two steps (reconstruction at distances $z_1$ and $z_2$) are chosen such that $z=z_1+z_2$. Here, the magnification is controlled by the choice of the intermediate reconstruction distance $z_1$. An optimization of this approach allows the authors of Ref.~\cite{Wang2008} to better match physical diffraction, thus obtaining high-fidelity reconstruction of magnified holograms. Control of reconstruction magnification, shift, and aberration compensation has also been proposed and realized using a digital lens, with adjustable parameters, in the reconstruction process~\cite{Colomb2006}.
				
				In the following subsection, we will focus on two algorithms allowing the adjustment of magnification in the reconstruction process and that are based either on the convolution~\cite{JCLi2011} or the 1-FFT~\cite{Restrepo2010} implementation of the Fresnel transform.

				\subsubsection{Digital quadratic lens method}
			This method is based on the convolution approach~\cite{JCLi2009,Picart2009}. Prior to reconstruction, the hologram is padded to the desired horizon and then multiplied by a digital spherical wavefront, acting as a quadratic lens, which is defined by:
			\begin{equation}
			\mathcal{L}\left(x\right)=\exp\left(-i\frac{\pi}{\lambda R_{c}}x^2\right),
			\label{eq:Quadlens}
			\end{equation}
			where $R_{c}$ denotes the curvature radius of $\mathcal{L}$. This curvature radius can be defined in terms of system magnification such that:
			\begin{equation}
			R_c=\frac{\gamma z}{\gamma-1}.
			\end{equation}
			Here, $\gamma$ is the ratio between the CCD horizon (of the padded hologram) and the object physical extent. Working with a spherical reconstruction wavefront modifies the physical reconstruction distance $z$ to $z'=\gamma z$. Thus, hologram reconstruction can be realized by computing the following relation:
			\begin{multline}
			 E_{\rm{rec}}\left(\xi\right)=\frac{\exp\left(\frac{i2\pi}{\lambda}z'\right)}{i\lambda z'}\\
			 \mathcal{F}^{-1}\left[\mathcal{F}\{E\left(x\right)\mathcal{L}\left(x\right)\}\mathcal{F}\left\{h_{z'}\left(x\right)\right\}\right],
			\end{multline}
			when working within a "three-FFT" scheme, or
			\begin{equation}
			 E_{\rm{rec}}\left(\xi\right)=\frac{\exp\left(\frac{i2\pi}{\lambda}z'\right)}{i\lambda z'}
			\mathcal{F}^{-1}\left[\mathcal{F}\{E\left(x\right)\mathcal{L}\left(x\right)\}H\left(u\right)\right],
			\end{equation}
			when angular spectrum propagation is considered.
			An alternative method, based on this formalism associated with a spatial filtering of the 1-FFT reconstructed hologram, allows the reconstruction of the local object field with an adjustable magnification~\cite{JCLi2011}. This method makes it possible to limit the effect of the reference beam distortions. However, one more FFT (two, when the angular spectrum implementation is considered) is needed to deal with the filtering step.
				
				\subsubsection{Fresnel-Bluestein transform}
				This approach is based on a ``clever'' expansion of Eq. (\ref{eq:1TF})~\cite{Restrepo2010}. In the kernel of the Fourier transform, the product $2np$ is rewritten as $2np=n^2+p^2-\left(p-n\right)^2$~\cite{Bluestein1970}, such that the discrete Fresnel transform can be expressed by:
			\begin{multline}
			E_{\rm{rec}}\left(p\right)=\frac{\exp\left(\frac{i2\pi}{\lambda}z\right)}{i\lambda z}\exp\left[-\frac{i\pi}{\lambda z}\Delta\xi\left(\Delta x-\Delta\xi\right)p^2\right]\\
			\times\sum_{n=0}^{N}E\left(n\right)\exp\left[\frac{i\pi}{\lambda z}\Delta x\left(\Delta x-\Delta\xi\right)n^2\right]\\
			\times\exp\left[\frac{i\pi}{\lambda z}\Delta x\Delta\xi\left(p-n\right)^2\right].
			\label{eq:Bluestein}
			\end{multline}
			  \begin{figure}[h]
				\centering
				\subfigure[]{\includegraphics*[width=3.4cm]{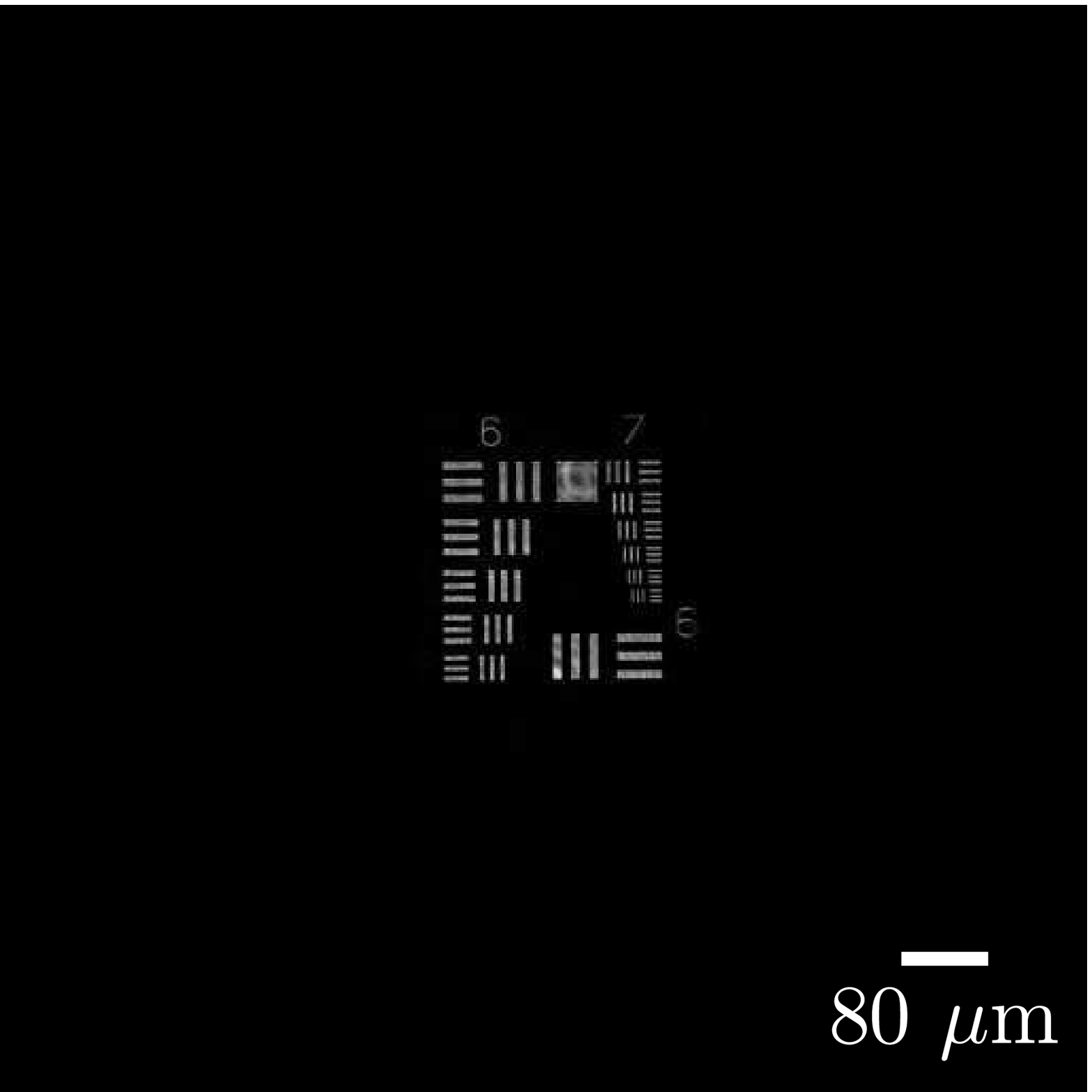}}
				\subfigure[]{\includegraphics*[width=3.4cm]{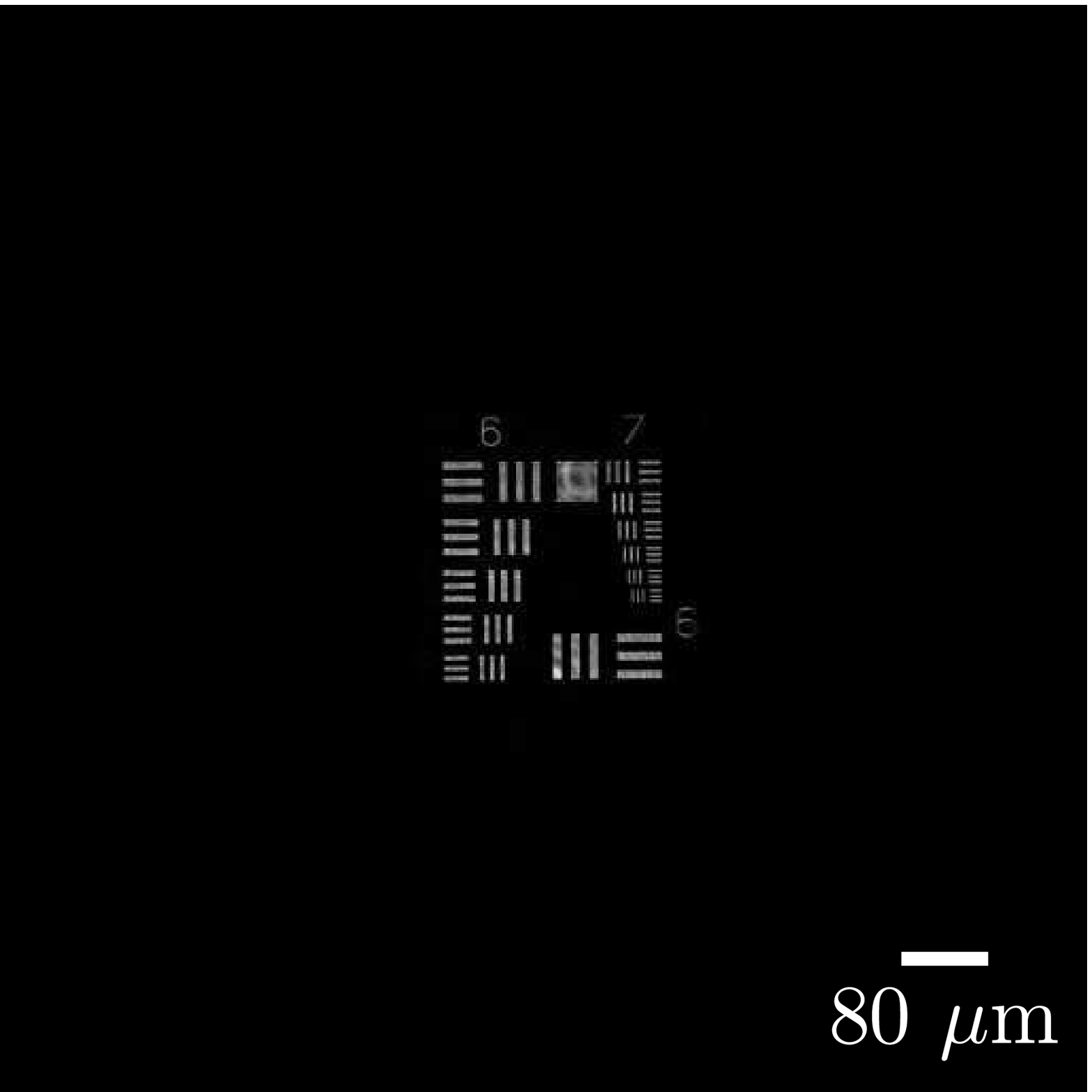}}
				\subfigure[]{\includegraphics*[width=3.4cm]{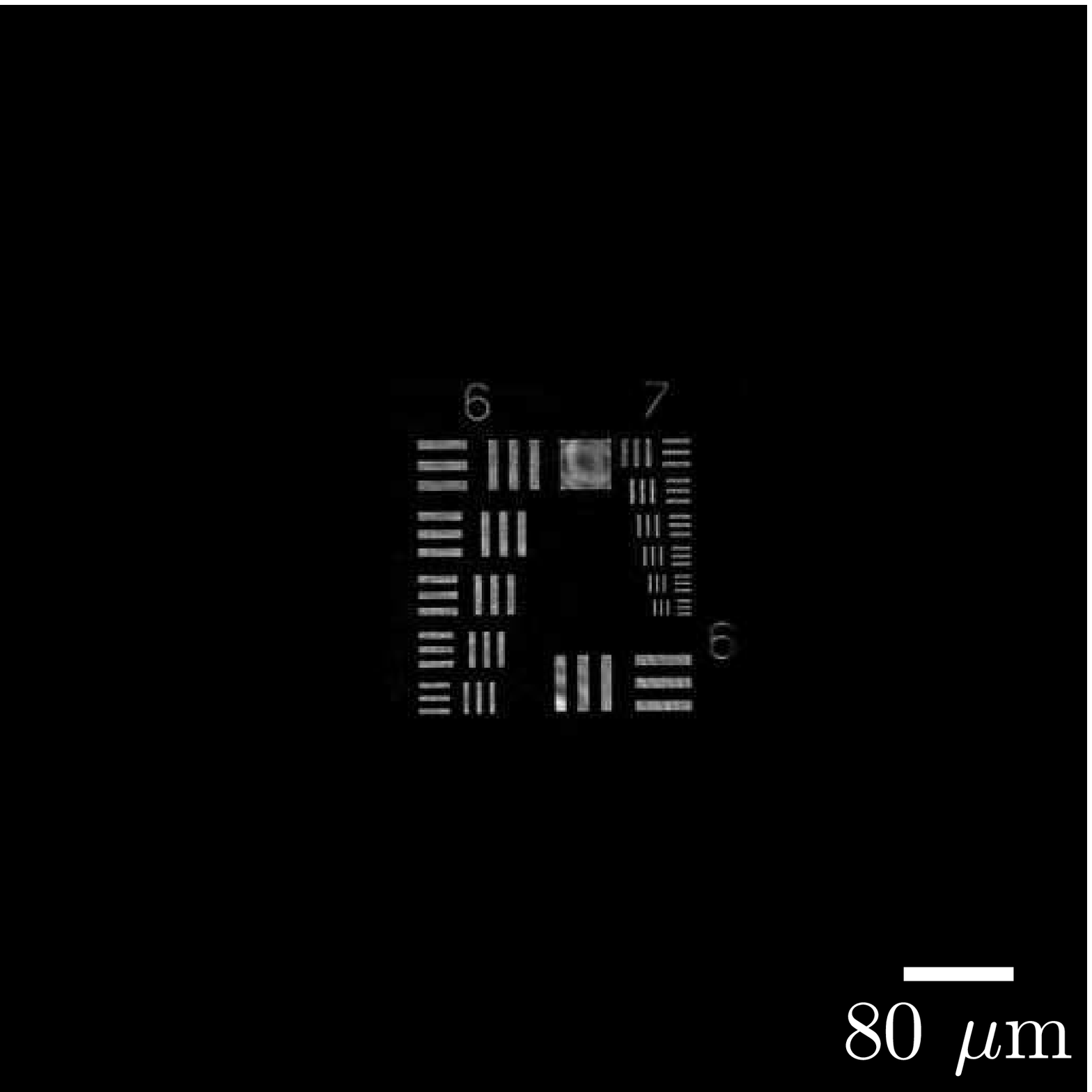}}
				\subfigure[]{\includegraphics*[width=3.4cm]{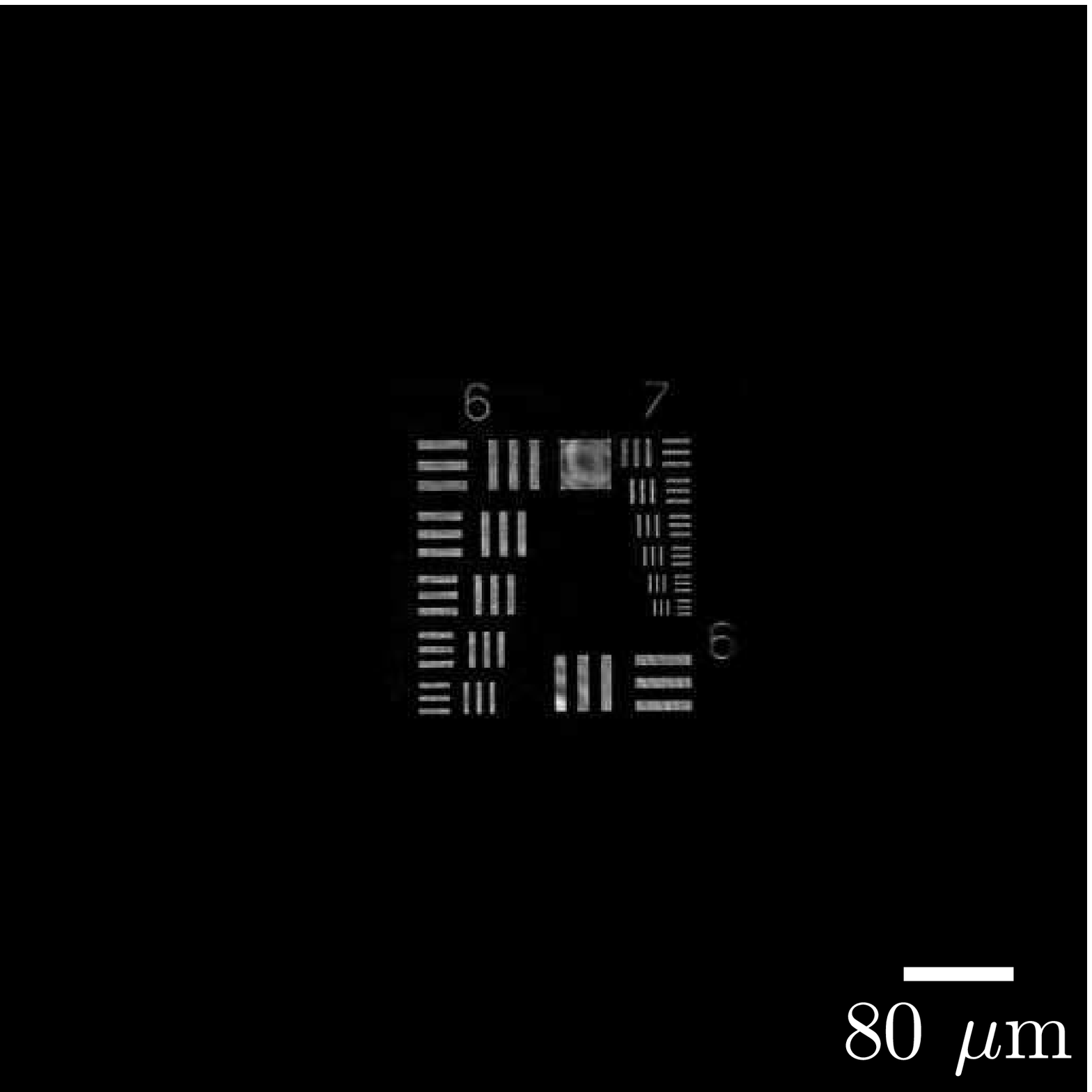}}
				\subfigure[]{\includegraphics*[width=3.4cm]{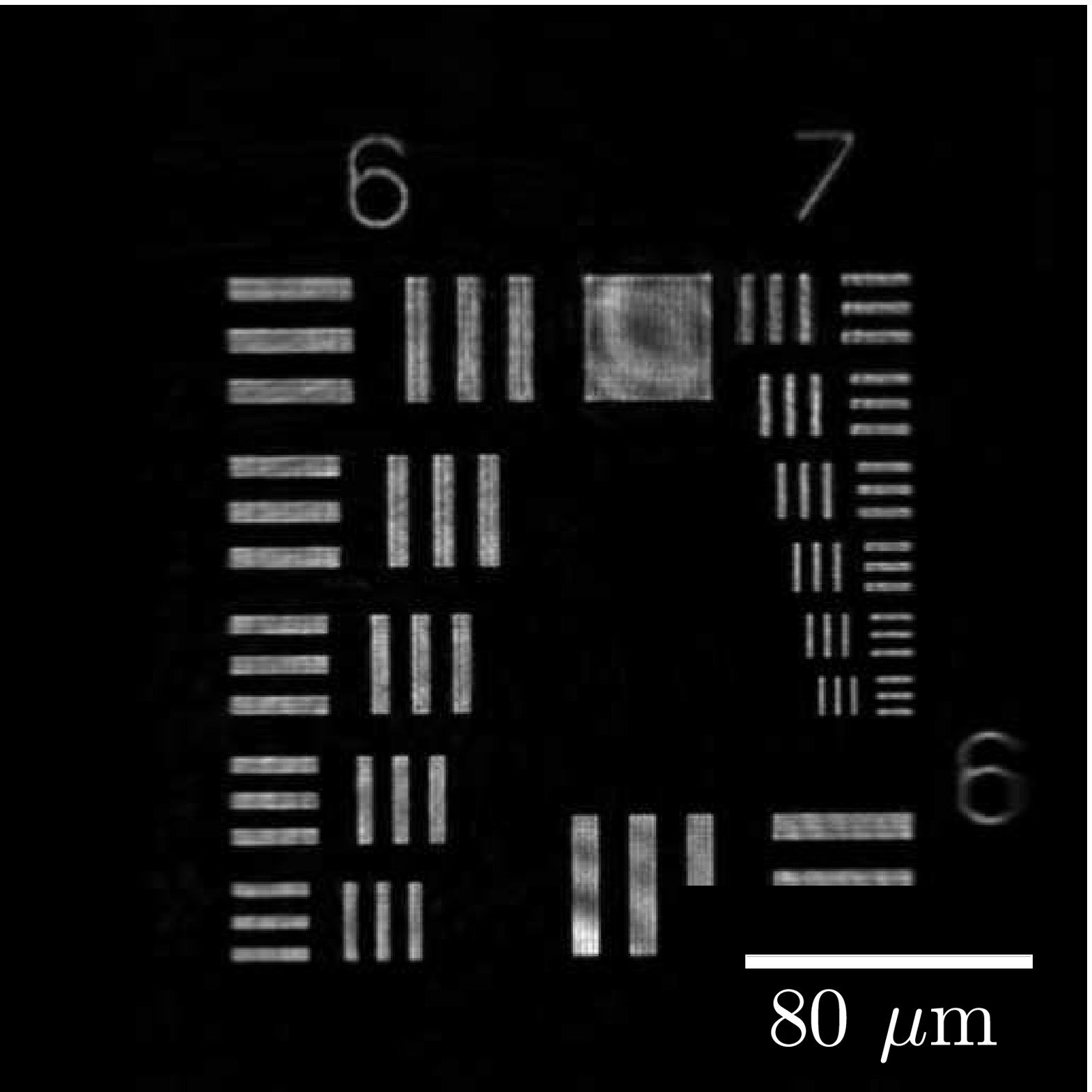}}
				\subfigure[]{\includegraphics*[width=3.4cm]{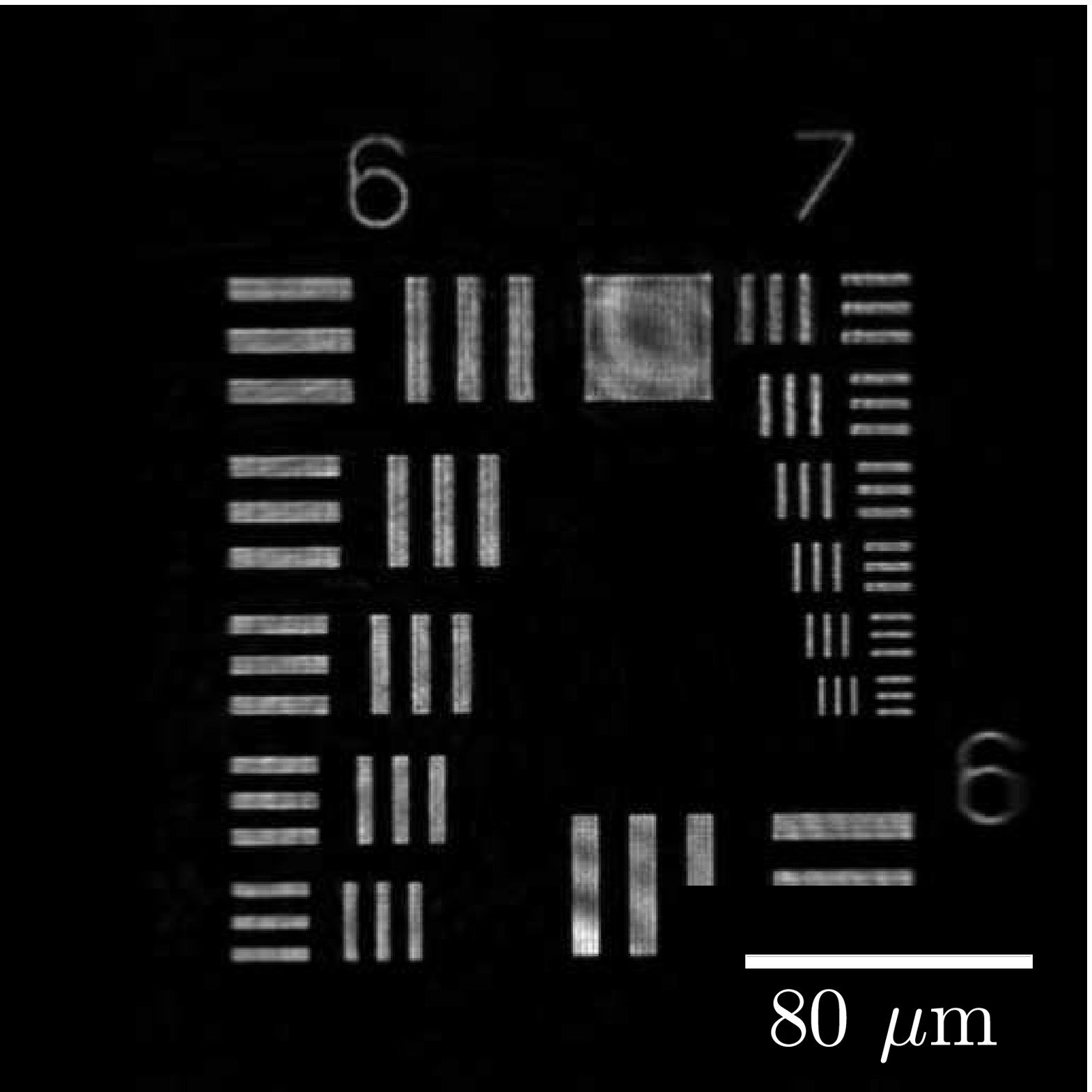}}
				\caption{Reconstruction of an hologram for $\gamma=0.8,1,2.5\times\gamma_0$. (a,c,e) Quadratic lens method. (b,d,f) Fresnel-Bluestein method.}\label{figpicgar}
				\end{figure}		
			Let $\gamma=\Delta\xi/\Delta x$ be the magnification of the reconstruction. Therefore, Eq. (\ref{eq:Bluestein}) can be rewritten as:
			\begin{multline}
			E_{\rm{rec}}\left(p\right)=\frac{\exp\left(\frac{i2\pi}{\lambda}z\right)}{i\lambda z}\exp\left[-\frac{i\pi}{\lambda z}\gamma\left(1-\gamma\right)\Delta x^2p^2\right]\\
			\times\sum_{n=0}^{N}E\left(n\right)\exp\left[\frac{i\pi}{\lambda z}\left(1-\gamma\right)n^2\Delta x^2\right]\\
			\times\exp\left[\frac{i\pi}{\lambda z}\gamma\left(p-n\right)^2\Delta x^2\right].
			\label{eq:Bluestein2}
			\end{multline}
			It should be noted that the magnification $\gamma$ is independent of the hologram recording parameters, and can be adjusted at will. With this formulation, Eq. (\ref{eq:Bluestein2}) is the spatial convolution product of two functions $f$ and $g$ defined by:
			\begin{equation}
			f(n)=E\left(n\right)\exp\left[i\frac{\pi}{\lambda z}\left(1-\gamma\right)n^2\Delta x^2\right],
			\end{equation}
			and
			\begin{equation}
			g(n)=\exp\left(i\frac{\pi}{\lambda z}\gamma n^2\Delta x^2\right).
			\end{equation}
			The Fresnel-Bluestein reconstruction algorithm can, therefore, be summarized as:
			\begin{multline}
			E_{\rm{rec}}\left(\xi\right)=\frac{\exp\left(\frac{i2\pi}{\lambda}z\right)}{i\lambda z}\exp\left[-\frac{i\pi}{\lambda z}\gamma\left(1-\gamma\right)\Delta x^2p^2\right]\\
  			\times\mathcal{F}^{-1}\left[\mathcal{F}\{f\left(x\right)\}\mathcal{F}\{g\left(x\right)\}\right].
			\end{multline}

Adjustable magnification rendering with the quadratic lens method and the Fresnel-Bluestein algorithm yield to the same results. Let $\gamma_0$ be the intrinsic magnification of the 1-FFT reconstruction scheme. Hologram rendering of a USAF resolution target sector, with $228\ \rm{line\ pairs}.mm^{-1}$ spatial frequency at element (7-6), with both methods at magnification $\gamma=0.8,1,2.5\times\gamma_0$ is reported in Fig. (\ref{figpicgar}). Despite these two methods are based on different formalisms (single-FFT formalism, and convolution based approach), it is here made obvious that both methods lead to the same results.

				\begin{figure*}[t]
				\centering
				\includegraphics[width=13cm]{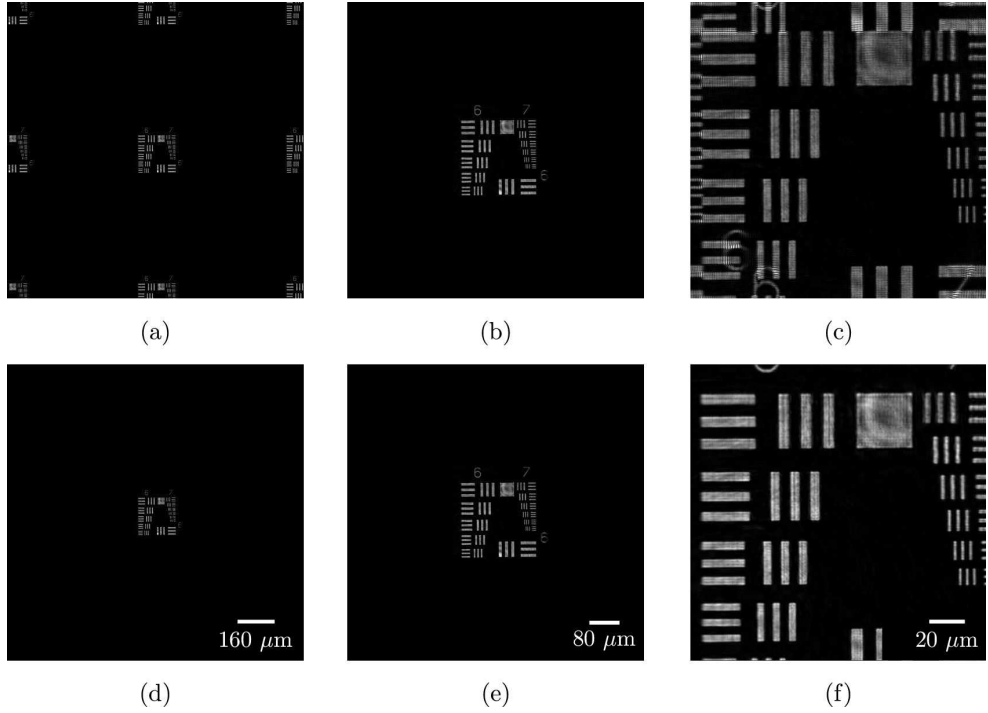}
				\caption{Illustration of alias and replica phenomena. (a) Reconstruction
				with $\gamma=0.5\times\gamma_0$. (c) Reconstruction
				with $\gamma=4\times\gamma_0$. (d) Same as (a) with replica removal. (f) Same as (c) with aliases filtering. (b,e) Reconstruction
				with $\gamma=\gamma_0$.}\label{fig4}
				\end{figure*}
				

				\subsubsection{Aliases and replicas}
			Adjustable magnification algorithms make it possible either to zoom over details in reconstructed images or to reconstruct objects whose dimensions are greater than that of the recording device. However, cares are to be taken. As a matter of fact, working with reconstruction horizons smaller than the object physical extend may cause aliases in the reconstruction plane, whereas replicas may appear in the opposite situation. In other words, considering $\gamma_0=\lambda z / \left(N \Delta x^2\right)$ the intrinsic magnification of the 1-FFT based Fresnel transform implementation,
			\begin{equation}
			\gamma<\gamma_0
			\end{equation}
			will lead to replicas in the reconstructed image, whereas choosing
			\begin{equation}
			\gamma>\gamma_0
			\end{equation}
			will generate aliases.
			
			\begin{figure}[!h]
				\centering
				\subfigure[]{\includegraphics*[width=7cm]{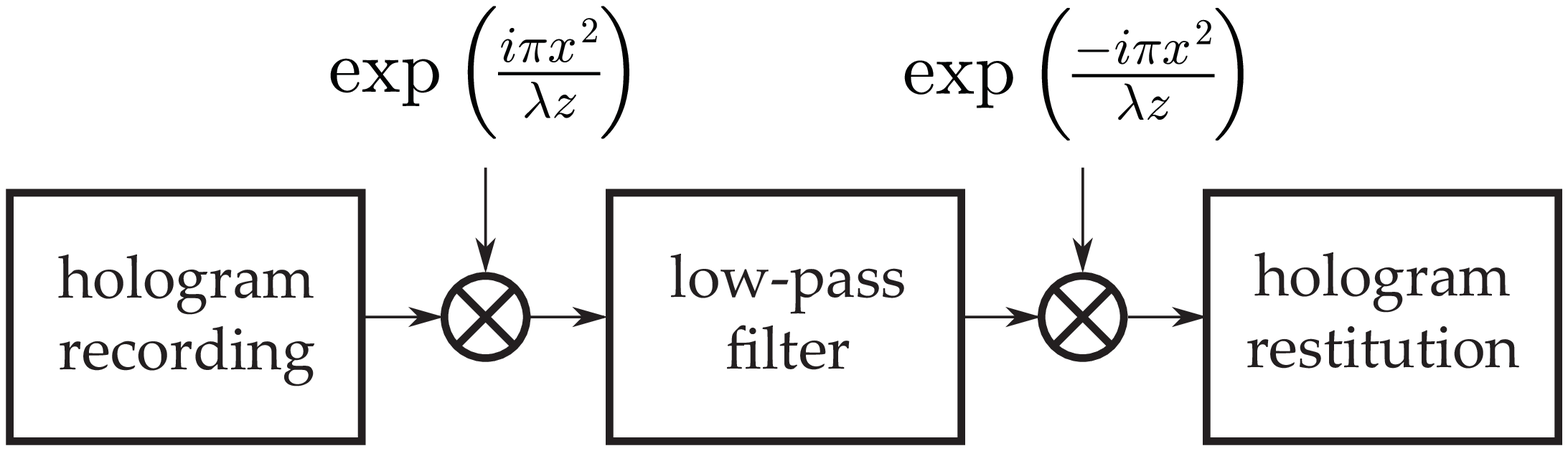}}
				\subfigure[]{\includegraphics*[width=7cm]{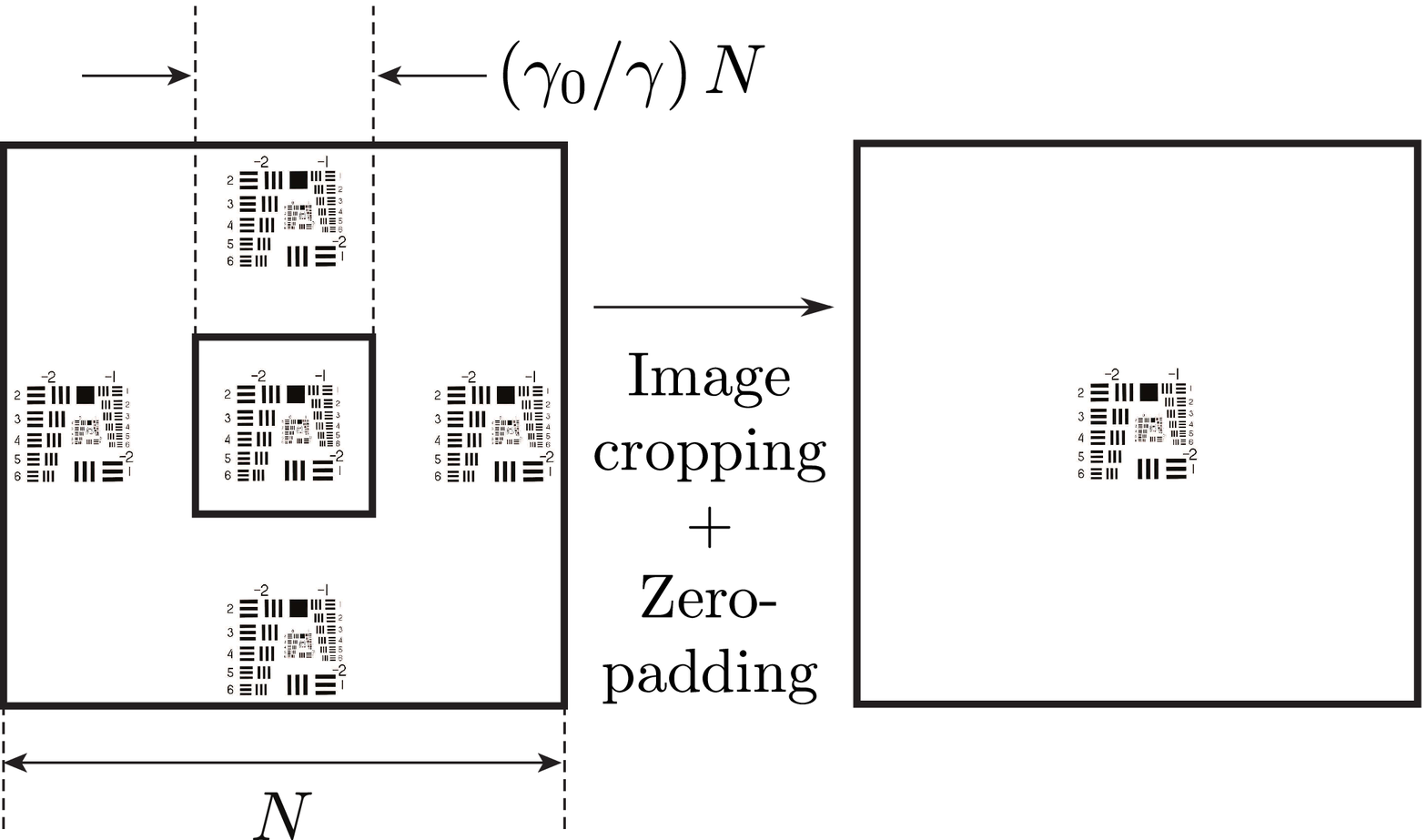}}
				\caption{(a) Synoptics of the anti-alias procedure. (b) Replica removal scheme.}\label{fig5}
			\end{figure}

			This aspect is illustrated by Fig. \ref{fig4} (a) and \ref{fig4} (c). A hologram of the object was reconstructed using the adjustable magnification algorithm proposed by Restrepo~\cite{Restrepo2010} (results would have been the same if the quadratic lens algorithm was considered). The reconstruction with the 1-FFT algorithm ($\gamma=\gamma_0$) is proposed on Fig. (\ref{fig4}) (b) and (e). It should be noted that replicas can be seen in Fig. \ref{fig4} (a) (here $\gamma<\gamma_0$) and aliases appear in Fig. \ref{fig4} (c) (for $\gamma>\gamma_0$). These unwanted effects degrade the reconstruction quality and must be avoided.												
				
				To limit the aliasing effect, when $\gamma>\gamma_0$, Hennelly proposed a filtering scheme which is presented in Fig. (\ref{fig5}) (a)~\cite{Hennelly2010}. Prior to being reconstructed, the hologram is multiplied by a chirp function defined as:
			\begin{equation}
			\mathcal{C}\left(x\right)=\exp\left(i\frac{\pi}{\lambda z}x^2\right).
			\end{equation}
			The resulting chirp-multiplied hologram is then low-pass filtered and finally multiplied by $\mathcal{C}^*$, which is the complex conjugate of $\mathcal{C}$. The size of the filtering window is chosen so as to match the physical extent of the reconstruction horizon at distance $z$. It is therefore possible to reconstruct the hologram using an adjustable magnification algorithm. Benefits of this filtering approach is illustrated by Figs. \ref{fig4} (c) and \ref{fig4} (f). Aliases artifacts are completely removed, thus giving a high contrast image of the reconstructed objects.
				
				When $\gamma<\gamma_0$, replicas can be seen on the reconstructed image. Their removal can be performed by the procedure illustrated in Fig. (\ref{fig5}) (b). The reconstructed hologram is cropped in order to keep the $\left(\gamma_0/\gamma\right)N$ pixels associated with the original object. This selection is then zero-padded to the original size of the hologram. As can be seen from Figs. \ref{fig4} (a) and \ref{fig4} (d), replicas are completely removed.

				\subsection{Fresnelet decomposition}
				\label{Sub:fresnelets}
			Fresnelet decomposition was initially proposed by Liebling for the reconstruction and processing of digital holograms~\cite{Liebling2003a}. This multiresolution scheme finds application in a wide variety of domains such as data compression~\cite{Darakis2006,Darakis2007}, non linear filtering~\cite{Liebling2003b}, wavefront retrieving~\cite{Liebling2004a} and can be considered in autofocusing procedures~\cite{Liebling2004b}. Fresnelet reconstruction of a hologram consists of its decomposition on a basis of Fresnel-transformed wavelets.
				
				Liebling proposed the use of B-splines, which can be defined as~\cite{Liebling2003a}:
			\begin{equation}
			 \beta^n\left(x\right)=\underbrace{\beta^0*\ldots*\beta^0}_{\rm{n+1}}\left(x\right),
			\end{equation}
			where $\beta^0$ is given by:
			\begin{equation}
			\beta^0\left(x\right)=\left\{
			\begin{array}{ll}
			1, & 0<x<1\\
			\frac{1}{2}, & x=0\ \rm{or}\ x=1\\
			0, & \rm{otherwise}
			\end{array}
			\right.
			,
			\end{equation}
			and the $*$ symbol denotes the convolution product.
			As shown by Unser, B-spline fulfills all the mathematical requirements to be used for multiresolution analysis of $L_2\left(\mathbb{R}\right)$~\cite{Unser1992}, especially the two-scale relation:
			\begin{equation}
			 \beta^n\left(\frac{x}{2}\right)=\sum_{k\in\mathbb{Z}}h\left(k\right)\beta^n\left(x-k\right).
			\end{equation}
			Here $h\left(k\right)=\frac{1}{2^n}\left(\begin{array}{c}n+1\\ k\end{array}\right)$ is the binomial filter.	
				
				\begin{figure}[h]
				\centering
				\includegraphics*[width=7cm]{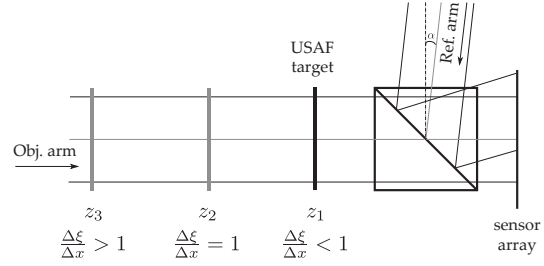}
				\caption{Experimental procedure for the holographic reconstruction benchmarking}\label{fig_bench}
				\end{figure}					
				
				B-spline can be used to generate a semi-orthogonal wavelet function basis of $L_2\left(\mathbb{R}\right)$ denoted $\psi_{j,k}^n$ and defined as:
			\begin{equation}
			 \left\{\psi_{j,k}^n=2^\frac{-j}{2}\psi^n\left(2^{-j}x-k\right)\right\}_{j,k\in\mathbb{Z}},
			\end{equation}
			where,
			\begin{equation}
			 \psi^n\left(\frac{x}{2}\right)=\sum_{k\in\mathbb{Z}}g\left(k\right)\beta^n\left(x-k\right).
			\end{equation}
			The filter $g\left(k\right)$ is the quadrature mirror filter of $h\left(k\right)$. Fresnelets basis can be calculated simply by taking the Fresnel transform of the B-spline basis. Fresnelet bases are therefore defined by:
			\begin{equation}
			 \left\{\tilde{\psi}_{j,k}^n=2^\frac{-j}{2}\tilde{\psi}^n\left(2^{-j}x-k\right)\right\}_{j,k\in\mathbb{Z}},
			\end{equation}
			with
			\begin{equation}
			 \tilde{\psi}^n\left(\frac{x}{2}\right)=\sum_{k\in\mathbb{Z}}g\left(k\right)\tilde{\beta}^n\left(x-k\right),
			 \label{eq:psi}
			\end{equation}
			and
			\begin{equation}
			 \tilde{\beta}^n\left(\frac{x}{2}\right)=\sum_{k\in\mathbb{Z}}h\left(k\right)\tilde{\beta}^n\left(x-k\right).
			 \label{eq:beta}
			\end{equation}
			Here $\tilde{.}$ is associated with the Fresnel transform.
				\begin{figure*}[t]
				\centering
				\includegraphics*[width=12cm]{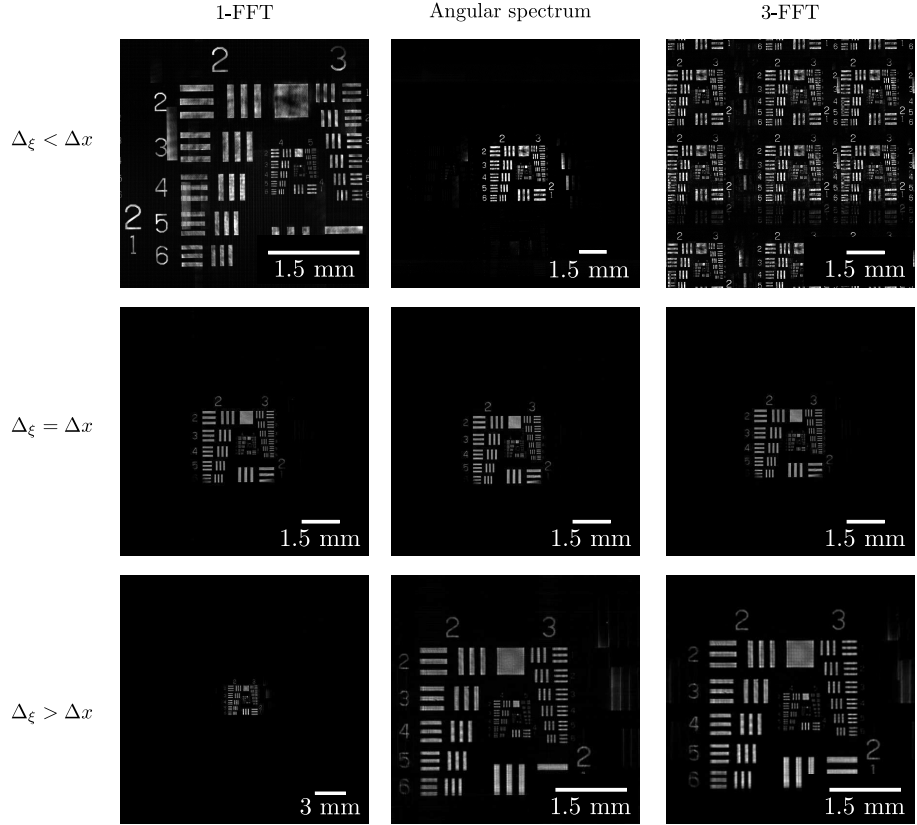}
				\caption{Holographic reconstructions of USAF target located at different distances.}\label{benchmark}
				\end{figure*}
				It should be noted that computation method chosen for the Fresnel transform will affect the Fresnelet transform results. As a matter of fact, Fresnelet transform properties will be the same as those of the chosen Fresnel computation scheme (\emph{e.g.} adjustable, or unitary magnification).

				\section{Application}
				\label{sec3}
			In this part, reconstruction of experimental holograms is performed according to the methods reported in Section (\ref{sec2}). Holograms are recorded according to the experimental set-up of Fig. (\ref{fig_bench}). Here, off-axis interference between the reference and object beam are recorded on a $2048\times2048$ pixels CCD sensor with  $\Delta x=7.4\ \mu \rm{m}$ pixel pitch. The object consists of an inverted USAF target illuminated with a green laser ($\lambda=532\ \rm{nm}$). The USAF target is positioned at three different distances $z_i$, from the sensor, chosen such that $\Delta \xi<\Delta x$, $\Delta \xi=\Delta x$, and $\Delta \xi>\Delta x$, where $\Delta \xi$ and $\Delta x$ denote the size of the reconstruction and of the CCD sensor respectively. Experimental reconstruction are presented hereafter.
					
			\subsection{Classical reconstruction methods}
							In this section, holograms recorded at $z_i$ such that $\Delta \xi<\Delta x$, $\Delta \xi=\Delta x$, and $\Delta \xi>\Delta x$, are reconstructed using 1-FFT, angular spectrum propagation, and 3-FFT methods. Images of the reconstructed objects are proposed in Fig. (\ref{benchmark}). This figure consists of a two entry table. In each row, the hologram reconstructed when $\Delta\xi<\Delta x$, $\Delta\xi=\Delta x$, and $\Delta\xi>\Delta x$ are depicted. Each column is associated with the chosen reconstruction method: 1-FFT, angular spectrum propagation, and 3-FFT. It is noticeable that in most cases reconstruction result depend on the method chosen. In the following section, results obtained are detailed row by row.
			
			\subsubsection{Reconstruction for $\Delta \xi<\Delta x$}
				\begin{enumerate}
				\item{1-FFT\\}
				As seen on Fig. (\ref{benchmark}), aliases are present in the reconstructed image of the object. This is due to the fact that the 1-FFT implementation of the Fresnel transform results in a magnified image of the original object. Thus, the reconstructed object extend over the limits of the CCD sensor.
				\item{Angular spectrum\\}
				As far as $\Delta \xi<\Delta x$, the reconstructed object is well embedded within the CCD sensor horizon. Angular spectrum method is therefore well suited for reconstruction of holograms recorded near the CCD sensor.
				\item{3-FFT\\}
				The reconstructed image of the object is embedded within the CCD sensor. However, replicas can be noticed from the reconstructed hologram. This is due to the fact that, when the reconstruction distance $z<N\Delta x/\lambda$, the impulse response $h_z$ is ill-sampled: in this situation, the sampling theorem is not verified~\cite{Onural2000}.
				\end{enumerate}				
				
							\subsubsection{Reconstruction for $\Delta \xi=\Delta x$}
				It can be noted that, for $\Delta \xi=\Delta x$, the three reconstruction methods considered give the same results. As a matter of fact, in this situation, the reconstructed horizon perfectly matches the sensor array extend. In other words, intrinsic magnification of 1-FFT Fresnel implementation is the same as that of convolution approaches.
			
\subsubsection{Reconstruction for $\Delta \xi>\Delta x$}
				\begin{enumerate}
				\item{1-FFT\\}
				As far as the object extent is bigger than the sensor array dimensions, the 1-FFT implementation of the Fresnel transform is appropriate for hologram reconstruction.
				\item{Angular spectrum and 3-FFT\\}
				The fact that these approaches exhibit unitary magnification is limiting when dealing with an object located far from the sensor. As a matter of fact, it can be realized from Fig. (\ref{benchmark}) that aliases occur in the reconstructed image of the hologram.
				\end{enumerate}
			
				\begin{figure}[h]
				\centering
				\includegraphics*[width=7cm]{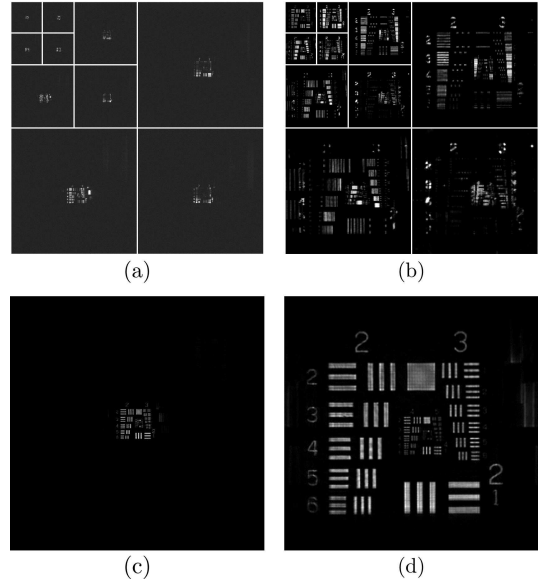}
				\caption{Fresnelet decomposition of the hologram recorded for $\Delta \xi>\Delta x$. (a) Fresnelet coefficients computed within the 1-FFT scheme. (b) Fresnelet coefficients computed within the 3-FFT scheme. (c) Hologram reconstruction from (a). (d) Hologram reconstruction from (b).}
				\label{Fresnelets}
				\end{figure}				
			
			Classical reconstruction methods have been applied to experimental holograms recorded at various distances from the sensor array. It can be noted that each method is valid only within a limited range of distances. In the next section, we will give a few word about Fresnelet decomposition, and show that its reconstruction properties can be modified to match each reconstruction method.

				\subsection{Fresnelets}
				As presented in Section \ref{Sub:fresnelets}, Fresnelet decomposition is similar to a multiscale-wavelet decomposition on a Fresnel-transformed base. One appealing feature of this decomposition is that the result of the Fresnelet reconstruction depends on the method chosen to compute the Fresnel transform of the wavelet base.
				
				To illustrate this point, Fresnelet reconstruction of the hologram recorded for $\Delta \xi>\Delta x$ is performed. In this situation, the 1-FFT method gave good results, whereas the 3-FFT reconstruction produces alises. Here, the fresnelet bases are calculated with the 1-FFT and the 3-FFT method according to Eqs. (\ref{eq:psi}) and (\ref{eq:beta}). Decompositions of the test hologram on the two calculated fresnelet bases are proposed in Fig. (\ref{Fresnelets}) (a) and (b) respectively. It should be noticed that the computation scheme chosen strongly affects the calculated coefficients. Therefore, properties of the fresnelet decomposition reconstruction depends on the method chosen to calculate the fresnelet base functions. This aspect is pointed out by Fig. (\ref{Fresnelets}) (c) and (d). Here, hologram reconstruction from the Fresnelet coefficients depicted in Fig. (\ref{Fresnelets}) (a) and (b) is realized. These reconstructions are similar to the one obtained with classical methods (See $\Delta \xi>\Delta x$ in Fig. (\ref{benchmark}) for comparison). Thus, for $\Delta \xi>\Delta x$, the single-FFT method will be more reliable than the 3-FFT scheme.
		
				\section{Conclusion}
				\label{conc}
				
We have proposed an overview of holographic reconstruction methods with application to off-axis intensity hologram treatment. Intrinsic properties and limitations of classical methods have been investigated and applicability ranges have been stated with the reconstruction of experimental holograms. It can be noted that the choice of the reconstruction method will be driven by the hologram recording conditions. For instance, when dealing with far and extended objects, 1-FFT algorithm will be the most appropriate, whereas convolution approaches will be suited for the reconstruction of small objects located near the sensor array. Adjustable magnification methods have been presented and can be viewed as a way to overcome limitations of the classical reconstruction schemes and allow the reconstruction result to be independent from the chosen scheme. Nevertheless, cares are to be taken in order to limit aliases and replicas when working with high or low magnification. Finally Fresnelet reconstruction of the holograms has been performed using the fact that the Fresnelet decomposition base depends on the method chosen to compute the Fresnel transform. This method can be considered for filtering or image compression when computational load is not a critical issue.

				\end{document}